\documentclass[italian,english]{article}
\usepackage[latin9]{inputenc}
\usepackage{color}
\usepackage{amssymb}

\makeatletter
\newcommand{\lyxaddress}[1]{
\par {\raggedright #1
\vspace{1.4em}
\noindent\par}
}

\makeatother

\usepackage{babel}

\begin{document}

\title{\textbf{An oscillating, homogeneous and isotropic Universe from Scalar-Tensor
gravity }}

\author{\textbf{Christian Corda}}

\maketitle

\lyxaddress{\begin{center}
Aerospace Tecnology Center, Av. Segle XXI s/n - Parc De La Marina,
E - 08840 Viladecans - Barcelona, Spain and Associazione Galileo Galilei,
Via Pier Cironi 16 - 59100 PRATO, Italy
\par\end{center}}

\lyxaddress{\begin{center}
\textit{E-mail address:} \textcolor{blue}{christian.corda@ego-gw.it}
\par\end{center}}

\begin{abstract}
An oscillating, homogeneous and isotropic Universe which arises from
Scalar-Tensor gravity is discussed in the linearized approach, showing
that some observative evidences like the Hubble Law and the Cosmological
Redshift are in agreement with the model. In this context Dark Energy
appears like a pure curvature effect arising by the scalar field.
\end{abstract}

\lyxaddress{PACS numbers: 04.80.Nn, 04.30.Nk, 04.50.+h}

\section{Introduction}

The accelerated expansion of the Universe, which is today observed,
shows that cosmological dynamic is dominated by the so called Dark
Energy which gives a large negative pressure. This is the standard
picture, in which such new ingredient is considered as a source of
the \textit{rhs} of the field equations. It should be some form of
un-clustered non-zero vacuum energy which, together with the clustered
Dark Matter, drives the global dynamics. This is the so called {}``concordance
model'' ($\Lambda CDM$) which gives, in agreement with the CMBR,
LSS and SNeIa data, a good trapestry of the today observed Universe,
but presents several shortcomings as the well known {}``coincidence''
and {}``cosmological constant'' problems \cite{key-1}. 

An alternative approach is changing the \textit{lhs} of the field
equations, seeing if observed cosmic dynamics can be achieved extending
General Relativity \cite{key-2,key-3,key-4}. In this different context,
it is not required to find out candidates for Dark Energy and Dark
Matter, that, till now, have not been found, but only the {}``observed''
ingredients, which are curvature and baryonic matter, have to be taken
into account. Considering this point of view, one can think that gravity
is not scale-invariant \cite{key-5} and a room for alternative theories
is present \cite{key-6,key-7,key-8}. In principle, the most popular
Dark Energy and Dark Matter models can be achieved considering $f(R)$
theories of gravity \cite{key-5,key-9}, where $R$ is the Ricci curvature
scalar. $f(R)$ theories of gravity are also conformally coupled with
Scalar-Tensor theories of gravity \cite{key-3,key-4,key-5,key-6,key-7,key-9}.
In this picture even the sensitive detectors for gravitational waves,
like bars and interferometers (i.e. those which are currently in operation
and the ones which are in a phase of planning and proposal stages)
\cite{key-10,key-11}, could, in principle, be important to confirm
or ruling out the physical consistency of General Relativity or of
any other theory of gravitation. This is because, in the context of
Extended Theories of Gravity, some differences between General Relativity
and the others theories can be pointed out starting by the linearized
theory of gravity \cite{key-12,key-13,key-14,key-15}. 

This paper is an integration of my previous research on oscillating
Universes \cite{key-7}. In \cite{key-7} it as been shown that an
oscillating, homogeneous and isotropic Universe which arises from
the linearized \textbf{$R^{2}$} theory of gravity is fine-tuned with
some observative evidences like the cosmological redshift and the
Hubble law. In this paper an oscillating, homogeneous and isotropic
Universe which arises by the linearized Scalar-Tensor gravity, recently
analysed from the point of view of gravitational waves in \cite{key-12,key-15},
is discussed, showing that the same observative evidences, i.e. the
Hubble Law and the Cosmological Redshift, are in agreement with the
model in this case too. In this context Dark Energy appears like a
pure curvature effect arising by the scalar field. The paper is organized
in this way: in the second section the linearization of Scalar-Tensor
gravity is performed, showing that a third mode of gravitational radiation
arises from the action \begin{equation}
S=\int d^{4}x\sqrt{-g}[f(\phi)R+\frac{1}{2}g^{\mu\nu}\phi_{;\mu}\phi_{;\nu}-V(\phi)+\mathcal{L}_{m}],\label{eq: scalar-tensor}\end{equation}

which is the Scalar-Tensor modification with respect the well known
canonical one of General Relativity (the Einstein - Hilbert action
\cite{key-16,key-17}), i.e.

\begin{equation}
S=\int d^{4}x\sqrt{-g}R+\mathcal{L}_{m}.\label{eq: EH}\end{equation}

In the third section, with the assumption that this third mode becomes
dominant at cosmological scales, following the linees of \cite{key-7}
an oscillating model of Universe will be shown.

In the fourth and fifth sections, the tuning with the Hubble Law and
the Cosmologic Redshift will be analysed.

\section{The linearized Scalar-Tensor gravity}

Choosing

\begin{equation}
\begin{array}{ccc}
\varphi=f(\phi) & \omega(\varphi)=\frac{f(\phi)}{2'f(\phi)} & W(\varphi)=V(\phi(\varphi))\end{array}\label{eq: scelta}\end{equation}

eq. (\ref{eq: scalar-tensor}) reads

\begin{equation}
S=\int d^{4}x\sqrt{-g}[\varphi R-\frac{\omega(\varphi)}{\varphi}g^{\mu\nu}\varphi_{;\mu}\varphi_{;\nu}-W(\varphi)+\mathcal{L}_{m}],\label{eq: scalar-tensor2}\end{equation}

which is a generalization of the Brans-Dicke theory \cite{key-23}.

By varying the action (\ref{eq: scalar-tensor2}) with respect to
$g_{\mu\nu}$ and the scalar field $\varphi,$ the field equations
are obtained (i.e. in this paper we work with $G=1$, $c=1$ and $\hbar=1$)
\cite{key-6,key-12}:

\begin{equation}
\begin{array}{c}
G_{\mu\nu}=-\frac{4\pi\tilde{G}}{\varphi}T_{\mu\nu}^{(m)}+\frac{\omega(\varphi)}{\varphi^{2}}(\varphi_{;\mu}\varphi_{;\nu}-\frac{1}{2}g_{\mu\nu}g^{\alpha\beta}\varphi_{;\alpha}\varphi_{;\beta})+\\
\\+\frac{1}{\varphi}(\varphi_{;\mu\nu}-g_{\mu\nu}\square\varphi)+\frac{1}{2\varphi}g_{\mu\nu}W(\varphi)\end{array}\label{eq: einstein-general}\end{equation}

with associed a Klein - Gordon equation for the scalar field

\begin{equation}
\square\varphi=\frac{1}{2\omega(\varphi)+3}(-4\pi\tilde{G}T^{(m)}+2W(\varphi)+\varphi W'(\varphi)+\frac{d\omega(\varphi)}{d\varphi}g^{\mu\nu}\varphi_{;\mu}\varphi_{;\nu}.\label{eq: KG}\end{equation}

In the above equations $T_{\mu\nu}^{(m)}$ is the ordinary stress-energy
tensor of the matter and $\tilde{G}$ is a dimensional, strictly positive,
constant \cite{key-6,key-12}. The Newton constant is replaced by
the effective coupling

\begin{equation}
G_{eff}=-\frac{1}{2\varphi},\label{eq: newton eff}\end{equation}

which is, in general, different from $G$. General Relativity is recovered
when the scalar field coupling is

\begin{equation}
\varphi=const=-\frac{1}{2}.\label{eq: varphi}\end{equation}

Because we want to study interactions at cosmological scales, the
linearized theory in vacuum ($T_{\mu\nu}^{(m)}=0$), which gives a
better approximation than Newtonian theory, can be analyzed, with
a little perturbation of the background, which is assumed given by
a Minkowskian background plus $\varphi=\varphi_{0}$. Because we are
limiting ourself to the linar approximation, $\omega$ can be considered
constant and fixed by $\varphi_{0}$ \cite{key-6,key-12}: 

\begin{equation}
\omega=\omega_{0}=\omega(\varphi_{0}).\label{eq: def omega zero}\end{equation}

$\varphi_{0}$ is also assumed to be a minimum for $W$: 

\begin{equation}
W\simeq\frac{1}{2}\alpha\delta\varphi^{2}\Rightarrow W'\simeq\alpha\delta\varphi.\label{eq: minimo}\end{equation}

Thus, putting

\begin{equation}
\begin{array}{c}
g_{\mu\nu}=\eta_{\mu\nu}+h_{\mu\nu}\\
\\\varphi=\varphi_{0}+\delta\varphi,\end{array}\label{eq: linearizza}\end{equation}

to first order in $h_{\mu\nu}$ and $\delta\varphi$, calling $\widetilde{R}_{\mu\nu\rho\sigma}$
, $\widetilde{R}_{\mu\nu}$ and $\widetilde{R}$ the linearized quantity
which correspond to $R_{\mu\nu\rho\sigma}$ , $R_{\mu\nu}$ and $R$,
the linearized field equations are obtained \cite{key-6,key-12}:

\begin{equation}
\begin{array}{c}
\widetilde{R}_{\mu\nu}-\frac{\widetilde{R}}{2}\eta_{\mu\nu}=\partial_{\mu}\partial_{\nu}\xi-\eta_{\mu\nu}\square\xi\\
\\{}\square\xi=-E^{2}\xi,\end{array}\label{eq: linearizzate1}\end{equation}

where

\begin{equation}
\begin{array}{c}
\xi\equiv\frac{\delta\varphi}{\varphi_{0}}\\
\\E^{2}\equiv\frac{\alpha\varphi_{0}}{2\omega_{0}+3}\end{array}\label{eq: definizione}\end{equation}

have been defined. $E$ represents the {}``curvature'' energy wich
arises from the scalar field.

$\widetilde{R}_{\mu\nu\rho\sigma}$ and eqs. (\ref{eq: linearizzate1})
are invariants for gauge transformations

\begin{equation}
\begin{array}{c}
h_{\mu\nu}\rightarrow h'_{\mu\nu}=h_{\mu\nu}-\partial_{(\mu}\epsilon_{\nu)}\\
\\\delta\varphi\rightarrow\delta\varphi'=\delta\varphi;\end{array}\label{eq: gauge}\end{equation}

then one can define

\begin{equation}
\bar{h}_{\mu\nu}\equiv h_{\mu\nu}-\frac{h}{2}\eta_{\mu\nu}-\eta_{\mu\nu}\xi\label{eq: ridefiniz}\end{equation}

and, considering the gauge transform (Lorenz condition) with the condition 

\begin{equation}
\square\epsilon_{\nu}=\partial^{\mu}\bar{h}_{\mu\nu}\label{eq:lorentziana}\end{equation}

for the parameter $\epsilon^{\mu}$:

\begin{equation}
\partial^{\mu}\bar{h}_{\mu\nu}=0,\label{eq: cond lorentz}\end{equation}

the field equations can be rewritten like

\begin{equation}
\square\bar{h}_{\mu\nu}=0\label{eq: onda T}\end{equation}

\begin{equation}
\square\xi=-E^{2}\xi;\label{eq: onda S}\end{equation}

solutions of eqs. (\ref{eq: onda T}) and (\ref{eq: onda S}) are
plan waves:

\begin{equation}
\bar{h}_{\mu\nu}=A_{\mu\nu}(\overrightarrow{p})\exp(ip^{\alpha}x_{\alpha})+c.c.\label{eq: sol T}\end{equation}

\begin{equation}
\xi=a(\overrightarrow{p})\exp(iq^{\alpha}x_{\alpha})+c.c.\label{eq: sol S}\end{equation}

where

\begin{equation}
\begin{array}{ccc}
k^{\alpha}\equiv(\omega,\overrightarrow{p}) &  & \omega=p\equiv|\overrightarrow{p}|\\
\\q^{\alpha}\equiv(\omega_{E},\overrightarrow{p}) &  & \omega_{E}=\sqrt{E^{2}+p^{2}}.\end{array}\label{eq: k e q}\end{equation}

In eqs. (\ref{eq: onda T}) and (\ref{eq: sol T}) the equation and
the solution for the tensorial waves exactly like in General Relativity
have been obtained \cite{key-17}, while eqs. (\ref{eq: onda S})
and (\ref{eq: sol S}) are respectively the equation and the solution
for the scalar mode.

Note: the dispersion law for the modes of the scalar field $\xi$
is not linear. The velocity of every tensorial mode $\bar{h}_{\mu\nu}$
is the light speed $c$, but the dispersion law (the second of eq.
(\ref{eq: k e q})) for the modes of $\xi$ is that of a massive field
which can be discussed like a wave-packet \cite{key-12,key-13}. Also,
the group-velocity of a wave-packet of $\xi$ centered in $\overrightarrow{p}$
is 

\begin{equation}
\overrightarrow{v_{G}}=\frac{\overrightarrow{p}}{\omega},\label{eq: velocita' di gruppo}\end{equation}

which is exactly the velocity of a massive particle with mass-energy
$E$ and momentum $\overrightarrow{p}$.

From the second of eqs. (\ref{eq: k e q}) and eq. (\ref{eq: velocita' di gruppo})
it is simple to obtain:

\begin{equation}
v_{G}=\frac{\sqrt{\omega^{2}-E^{2}}}{\omega}.\label{eq: velocita' di gruppo 2}\end{equation}

Then, wanting a constant speed of the wave-packet, one needs

\begin{equation}
E=\sqrt{(1-v_{G}^{2})}\omega.\label{eq: relazione massa-frequenza}\end{equation}

Now let us remain in the Lorenz gauge with trasformations of the type
$\square\epsilon_{\nu}=0$; this gauge gives a condition of transversality
for the tensorial part of the field: $k^{\mu}A_{\mu\nu}=0$, but we
do not know if the total field $h_{\mu\nu}$ is transverse. From eq.
(\ref{eq: ridefiniz}) it is

\begin{equation}
h_{\mu\nu}=\bar{h}_{\mu\nu}-\frac{\bar{h}}{2}\eta_{\mu\nu}-\eta_{\mu\nu}\xi.\label{eq: ridefiniz 2}\end{equation}

At this point, if being in the massless case one could put

\begin{equation}
\begin{array}{c}
\square\epsilon^{\mu}=0\\
\\\partial_{\mu}\epsilon^{\mu}=-\frac{\bar{h}}{2}-\xi,\end{array}\label{eq: gauge2}\end{equation}

which gives the total transversality of the field. But in the massive
case this is impossible. In fact, if applying the d' Alembertian operator
to the second of eqs. (\ref{eq: gauge2}) and using the field equations
(\ref{eq: onda T}) and (\ref{eq: onda S}) it is

\begin{equation}
\square\epsilon^{\mu}=-E^{2}\xi,\label{eq: contrasto}\end{equation}

which is in contrast with the first of eqs. (\ref{eq: gauge2}). In
the same way it is possible to show that it does not exist any linear
relation between the tensorial field $\bar{h}_{\mu\nu}$ and the scalar
field $\xi$. Thus, a gauge in wich $h_{\mu\nu}$ is purely spatial
cannot be chosen (i.e. we cannot put $h_{\mu0}=0,$ see eq. (\ref{eq: ridefiniz 2})).
But the traceless condition to the field $\bar{h}_{\mu\nu}$ can be
put:

\begin{equation}
\begin{array}{c}
\square\epsilon^{\mu}=0\\
\\\partial_{\mu}\epsilon^{\mu}=-\frac{\bar{h}}{2},\end{array}\label{eq: gauge traceless}\end{equation}

which implies

\begin{equation}
\partial^{\mu}\bar{h}_{\mu\nu}=0.\label{eq: vincolo}\end{equation}

Wanting to save the conditions $\partial_{\mu}\bar{h}^{\mu\nu}$ and
$\bar{h}=0$, transformations like

\begin{equation}
\begin{array}{c}
\square\epsilon^{\mu}=0\\
\\\partial_{\mu}\epsilon^{\mu}=0\end{array}\label{eq: gauge 3}\end{equation}

can be used and, taking $\overrightarrow{p}$ in the $z$ direction,
one can choose a gauge in which only $A_{11}$, $A_{22}$, and $A_{12}=A_{21}$
are different to zero. The condition $\bar{h}=0$ gives $A_{11}=-A_{22}$.
Now, putting these equations in eq. (\ref{eq: ridefiniz 2}) and defining
$\Phi\equiv-\xi$ one obtains

\begin{equation}
h_{\mu\nu}(t,z)=A^{+}(t-z)e_{\mu\nu}^{(+)}+A^{\times}(t-z)e_{\mu\nu}^{(\times)}+\Phi(t-v_{G}z)\eta_{\mu\nu}.\label{eq: perturbazione totale}\end{equation}

The term $A^{+}(t-z)e_{\mu\nu}^{(+)}+A^{\times}(t-z)e_{\mu\nu}^{(\times)}$
describes the two standard (i.e. tensorial) polarizations of gravitational
waves which arises from General Relativity, while the term $\Phi(t-v_{G}z)\eta_{\mu\nu}$
is the scalar field.

\section{An oscillating Cosmology}

Before starting the analysis, one has to emphasize that, in a cosmologic
framework, the linerar approximation has to be considered a good approximation.
This arises from two considerations. 

\begin{enumerate}
\item It is well known that, when treating cosmologic problems, astrophysicsts
often limit their analyses to Newtonian theory. The linearized approximation
represents the substitution of the Newtonian approximation with a
less restrictive hypothesis. In this case, even if the field remains
weak, it can vary with time and restrictions in the motion of test
particles are not present. New physics phenomena, in respect to Newtonian
approximation, like light deflection and gravitational radiation,
arises from this hypotesis. It is enlighting that, in the framework
of standard General Relativity, Einstein used the linearized approach
in the discussion about observative predictions of the theory \cite{key-17}.
\item As our observations are performed on Earth, the coordinate system
in which the space-time is locally flat has to be used and the distance
between any two points is given simply by the difference in their
coordinates in the sense of post-Newtonian physics. This is exactly
the sense of a spacetime which is considere \textit{globally} curved
but \textit{locally} flat \cite{key-17}.
\end{enumerate}
By assuming that, at cosmological scales, the third mode becomes dominant
(i.e. $A^{+},A^{-}\ll\Phi$) \cite{key-7}, as it appears from observations,
eq. (\ref{eq: perturbazione totale}) can be rewritten as

\begin{equation}
h_{\mu\nu}(t,z)=\Phi(t,z)\eta_{\mu\nu}\label{eq: perturbazione scalare}\end{equation}
and the corrispondent line element is the conformally flat one

\begin{equation}
ds^{2}=[1+\Phi(t,z)](-dt^{2}+dz^{2}+dx^{2}+dy^{2}).\label{eq: metrica puramente scalare}\end{equation}

Defining 

\begin{equation}
a^{2}\equiv1+\Phi(t,z),\label{eq: a quadro}\end{equation}

equation (\ref{eq: metrica puramente scalare}) becomes similiar to
the well known cosmological Friedmann- Robertson Walker (FRW) line
element of the standard homogeneus and isotropic flat Universe which
is well known in the literature \cite{key-16,key-17,key-18,key-19,key-20}:

\begin{equation}
ds^{2}=[a^{2}(t,z)](-dt^{2}+dz^{2}+dx^{2}+dy^{2}).\label{eq: metrica FRW}\end{equation}
In the linearized approach it is also \cite{key-7} \begin{equation}
a\simeq1+\frac{1}{2}\Phi(t,z),\label{eq: a}\end{equation}

which shows that in the model the scale factor of the Universe oscillates
near the (normalized) unity.

Below it will be shown that the model realizes an oscillating homogeneus
and isotropic Universe, but, before starting with the analysis, we
have to recall that observations today agrees with homogeneity and
isotropy.

In Cosmology, the Universe is seen like a dynamic and thermodynamic
system in which test masses (i.e. the {}``particles'') are the galaxies
that are stellar systems with a number of the order of $10^{9}-10^{11}$
stars \cite{key-16,key-17,key-18,key-19,key-20}. Galaxies are located
in clusters and super clusters, and observations show that, on cosmological
scales, their distribution is uniform. This is also confirmed by the
WMAP data on the Cosmic Background Radiation \cite{key-21,key-22}.
These assumption can be summarized in the so called Cosmological Principle:
\textit{the Universe is homogeneous everyway and isotropic around
every point.} Cosmological Principle semplifies the analysis of the
large scale structure, because it implies that the proper distances
between any two galaxies is given by an universal scale factor which
is the same for any couple of galaxies \cite{key-16,key-17,key-18,key-19,key-20}.

As our observations are performed in a laboratory environment on Earth,
the coordinate system in which the space-time is locally flat has
to be used and the distance between any two points is given simply
by the difference in their coordinates in the sense of Newtonian physics
\cite{key-12,key-13,key-14,key-15,key-16,key-17}. This frame is the
proper reference frame of a local observer, which we assume to be
located on Earth. In this frame gravitational signals manifest themself
by exerting tidal forces on the test masses, which are the galaxies
of the Universe. A detailed analysis of the frame of the local observer
is given in ref. \cite{key-17}, sect. 13.6. Here only the more important
features of this coordinate system are recalled:

the time coordinate $x_{0}$ is the proper time of the observer O;

spatial axes are centered in O;

in the special case of zero acceleration and zero rotation the spatial
coordinates $x_{j}$ are the proper distances along the axes and the
frame of the local observer reduces to a local Lorentz frame: in this
case the line element reads \cite{key-17}

\begin{equation}
ds^{2}=-(dx^{0})^{2}+\delta_{ij}dx^{i}dx^{j}+O(|x^{j}|^{2})dx^{\alpha}dx^{\beta}.\label{eq: metrica local lorentz}\end{equation}

The effect of the gravitational force on test masses is described
by the equation

\begin{equation}
\ddot{x^{i}}=-\widetilde{R}_{0k0}^{i}x^{k},\label{eq: deviazione geodetiche}\end{equation}
which is the equation for geodesic deviation in this frame.

Thus, to study the effect of the third mode of the linearized Scalar-Tensor
gravity on the galaxies, $\widetilde{R}_{0k0}^{i}$ has to be computed
in the proper reference frame of the Earth. But, because the linearized
Riemann tensor $\widetilde{R}_{\mu\nu\rho\sigma}$ is invariant under
gauge transformations \cite{key-12,key-13,key-17}, it can be directly
computed from eq. (\ref{eq: perturbazione scalare}). 

From \cite{key-17} it is:

\begin{equation}
\widetilde{R}_{\mu\nu\rho\sigma}=\frac{1}{2}\{\partial_{\mu}\partial_{\beta}h_{\alpha\nu}+\partial_{\nu}\partial_{\alpha}h_{\mu\beta}-\partial_{\alpha}\partial_{\beta}h_{\mu\nu}-\partial_{\mu}\partial_{\nu}h_{\alpha\beta}\},\label{eq: riemann lineare}\end{equation}

that, in the case eq. (\ref{eq: perturbazione scalare}), begins

\begin{equation}
\widetilde{R}_{0\gamma0}^{\alpha}=\frac{1}{2}\{\partial^{\alpha}\partial_{0}\Phi\eta_{0\gamma}+\partial_{0}\partial_{\gamma}\Phi\delta_{0}^{\alpha}-\partial^{\alpha}\partial_{\gamma}\Phi\eta_{00}-\partial_{0}\partial_{0}\Phi\delta_{\gamma}^{\alpha}\};\label{eq: riemann lin scalare}\end{equation}

the different elements are (only the non zero ones will be written):

\begin{equation}
\partial^{\alpha}\partial_{0}\Phi\eta_{0\gamma}=\left\{ \begin{array}{ccc}
\partial_{t}^{2}\Phi & for & \alpha=\gamma=0\\
\\-\partial_{z}\partial_{t}\Phi & for & \alpha=3;\gamma=0\end{array}\right\} \label{eq: calcoli}\end{equation}

\begin{equation}
\partial_{0}\partial_{\gamma}\Phi\delta_{0}^{\alpha}=\left\{ \begin{array}{ccc}
\partial_{t}^{2}\Phi & for & \alpha=\gamma=0\\
\\\partial_{t}\partial_{z}\Phi & for & \alpha=0;\gamma=3\end{array}\right\} \label{eq: calcoli2}\end{equation}

\begin{equation}
-\partial^{\alpha}\partial_{\gamma}\Phi\eta_{00}=\partial^{\alpha}\partial_{\gamma}\Phi=\left\{ \begin{array}{ccc}
-\partial_{t}^{2}\Phi & for & \alpha=\gamma=0\\
\\\partial_{z}^{2}\Phi & for & \alpha=\gamma=3\\
\\-\partial_{t}\partial_{z}\Phi & for & \alpha=0;\gamma=3\\
\\\partial_{z}\partial_{t}\Phi & for & \alpha=3;\gamma=0\end{array}\right\} \label{eq: calcoli3}\end{equation}

\begin{equation}
-\partial_{0}\partial_{0}\Phi\delta_{\gamma}^{\alpha}=\begin{array}{ccc}
-\partial_{t}^{2}\Phi & for & \alpha=\gamma\end{array}.\label{eq: calcoli4}\end{equation}

Now, putting these results in eq. (\ref{eq: riemann lin scalare}),
it results:

\begin{equation}
\begin{array}{c}
\widetilde{R}_{010}^{1}=-\frac{1}{2}\ddot{\Phi}\\
\\\widetilde{R}_{020}^{2}=-\frac{1}{2}\ddot{\Phi}\\
\\\widetilde{R}_{030}^{3}=\frac{1}{2}(\partial_{z}^{2}\Phi-\partial_{t}^{2}\Phi).\end{array}\label{eq: componenti riemann}\end{equation}

But, the assumption of homogenity and isotropy implies $\partial_{z}\Phi=0$
\cite{key-7}, which also implies

\begin{equation}
\begin{array}{c}
\widetilde{R}_{010}^{1}=-\frac{1}{2}\ddot{\Phi}\\
\\\widetilde{R}_{0120}^{2}=-\frac{1}{2}\ddot{\Phi}\\
\\\widetilde{R}_{030}^{3}=-\frac{1}{2}\ddot{\Phi}\end{array}\label{eq: componenti riemann 2}\end{equation}

which show that the oscillations of the Universe are the same in any
direction. 

Infact, using eq. (\ref{eq: deviazione geodetiche}), it results

\begin{equation}
\ddot{x}=\frac{1}{2}\ddot{\Phi}x,\label{eq: accelerazione mareale lungo x}\end{equation}

\begin{equation}
\ddot{y}=\frac{1}{2}\ddot{\Phi}y\label{eq: accelerazione mareale lungo y}\end{equation}

and 

\begin{equation}
\ddot{z}=\frac{1}{2}\ddot{\Phi}z,\label{eq: accelerazione mareale lungo z}\end{equation}

which are three perfectly symmetric oscillations.

\section{Consistence with observations: i) The Hubble Law}

The expansion of the Universe arises from the observations of E Hubble
in 1929 \cite{key-16,key-17,key-18,key-19,key-20}. The Hubble law
states that, galaxies which are at a distance $D,$ drift away from
Earth with a velocity 

\begin{equation}
v=H_{0}D.\label{eq: legge Hubble}\end{equation}

The today's Hubble expansion rate is 

\begin{equation}
H_{0}=h_{100}\frac{100Km}{sec\times Mpc}=3.2\times10^{-18}\frac{h_{100}}{sec}.\label{eq: cost. Hubble}\end{equation}

A dimensionless factor $h_{100}$ is included, which now is just a
convenience (in the past it came from an uncertainty in the value
of $H_{0}$). From the WMAP data it is $h_{100}=0.72\pm0.05$ \cite{key-21,key-22}.

Calling $f$ the frequency of the {}``cosmologic'' gravitational
wave and assuming that $f\ll H_{0}$ (i.e. the gravitational wave
is {}``frozen'' with respect the cosmological observations), the
observations of Hubble and the more recent ones imply that our model
of oscillating Universe has to be in the expansion phase \cite{key-7}

A good way to analyse proper distances (which are equal to proper
times in natural units) between two test masses is by means of light
rays \cite{key-12,key-13,key-17}. For the assumption of homogeneity
and isotropy, only the radial propagation of the light can be taken
into account.

In spherical coordinates equations (\ref{eq: accelerazione mareale lungo x}),
(\ref{eq: accelerazione mareale lungo y}) and (\ref{eq: accelerazione mareale lungo z})
give for the radial coordinate \begin{equation}
\ddot{r}=\frac{1}{2}\ddot{\Phi}r.\label{eq: accelerazione mareale lungo r}\end{equation}

Equivalently we can say that there is a gravitational potential \begin{equation}
V(\overrightarrow{r},t)=-\frac{1}{4}\ddot{\Phi}(t)r^{2},\label{eq:potenziale in gauge Lorentziana}\end{equation}

which generates the tidal forces, and that the motion of the test
masses is governed by the Newtonian equation \cite{key-7} 

\begin{equation}
\ddot{\overrightarrow{r}}=-\bigtriangledown V.\label{eq: Newtoniana}\end{equation}

Because we are in the linearized theory, following \cite{key-7},
the solution of eq. (\ref{eq: accelerazione mareale lungo r}) can
be found by using the perturbation method \cite{key-16,key-17}, obtaining

\begin{equation}
D=D_{0}+\frac{1}{2}D_{0}\Phi(t)=(1+\frac{1}{2}\Phi)D_{0}=a(t)D_{0}\label{eq:  distance}\end{equation}

Deriving this equation with respect the time we also get \begin{equation}
\frac{dD}{dt}=D_{0}\frac{da(t)}{dt}.\label{eq:  derivative}\end{equation}

Thus the Hubble law is obtained: \begin{equation}
\frac{1}{D}\frac{dD}{dt}=H_{0},\label{eq:  Hubble 2}\end{equation}

where 

\begin{equation}
H_{0}=(\frac{1}{a}\frac{da}{dt})_{0}.\label{eq:  const Hubble 2}\end{equation}

\section{Consistence with observations: ii) The Cosmological Redshift}

Let us now consider another point of view. The conformal line element
(\ref{eq: metrica FRW}) can be put in spherical coordinates, obtaining
\cite{key-7} \begin{equation}
ds^{2}=[1+\Phi(t)](-dt^{2}+dr^{2}).\label{eq: metrica puramente scalare radiale}\end{equation}

In this line element the angular coordinates have been neglected because
of the assumption of homogeneity and isotropy. The condition of null
geodesic in the above equation gives \begin{equation}
dt^{2}=dr^{2}.\label{eq: metrica puramente piu' di Corda lungo x 2}\end{equation}

Thus, from eq. (\ref{eq: metrica puramente piu' di Corda lungo x 2}),
it results that the coordinate velocity of the photon in the gauge
(\ref{eq: metrica puramente scalare radiale}) is equal to the speed
of light. This because in the coordinates (\ref{eq: metrica puramente scalare radiale})
$t$ is only a time coordinate. The rate $d\tau$ of the proper time
(distance) is related to the rate $dt$ of the time coordinate from
\cite{key-16}

\begin{equation}
d\tau^{2}=g_{00}dt^{2}.\label{eq: relazione temporale}\end{equation}
From eq. (\ref{eq: metrica puramente scalare radiale}) it is $g_{00}=(1+\Phi)$.
Then, using eq. (\ref{eq: metrica puramente piu' di Corda lungo x 2}),
we obtain

\begin{equation}
d\tau^{2}=(1+\Phi))dr^{2},\label{eq: relazione spazial-temporale}\end{equation}

which gives 

\begin{equation}
d\tau=\pm[(1+\Phi)]^{\frac{1}{2}}dr\simeq\pm[(1+\frac{1}{2}\Phi)]dr.\label{eq: relazione temporale 2}\end{equation}

We assume that photons are travelling by the galaxy to Earth in this
case too, thus the negative sign is needed \cite{key-7}. 

Integrating this equation it is\begin{equation}
\int_{\tau_{1}}^{\tau_{0}}\frac{d\tau}{1+\frac{1}{2}\Phi(t)}=\int_{r_{g}}^{0}dr=r_{g},\label{eq: redshift 1}\end{equation}

where $\tau_{1}$ and $\tau_{0}$ are the emission and reception istants
of the photon from galaxy and Earth respectively. If the light is
emitted with a delay $\bigtriangleup\tau_{1},$ it arrives on Earth
with a delay $\bigtriangleup\tau_{0}$. In this way \begin{equation}
\int_{\tau_{1}}^{\tau_{0}}\frac{d\tau}{1+\frac{1}{2}\Phi(t)}=\int_{\tau_{1}+\bigtriangleup\tau_{1}}^{\tau_{0}+\bigtriangleup\tau_{0}}\frac{d\tau}{1+\frac{1}{2}\Phi(t)}=r_{g}.\label{eq: redshift 2}\end{equation}

The radial coordinate $r_{g}$ is \textit{comoving} (i.e. constant
in the gauge (\ref{eq: metrica puramente scalare radiale})) because
the assumption of homogenity and isotropy implies $\partial_{z}\Phi=0,$
which removes the $z$ dependence in the line element (\ref{eq: metrica FRW}).
Thus the only dependence in the line element (\ref{eq: metrica puramente scalare radiale})
is the $t$ dependence in the scale factor $a=1+\frac{1}{2}\Phi(t)$.
Then, from equation (\ref{eq: redshift 2}) it is \begin{equation}
\begin{array}{c}
\int_{\tau_{1}}^{\tau_{0}}\frac{d\tau}{1+\frac{1}{2}\Phi(t)}=\int_{\tau_{1}}^{\tau_{0}}\frac{d\tau}{1+\frac{1}{2}\Phi(t)}+\\
\\+\int_{\tau_{1}}^{\tau_{0}+\bigtriangleup\tau_{0}}\frac{d\tau}{1+\frac{1}{2}\Phi(t)}-\int_{\tau_{0}}^{\tau_{1}+\bigtriangleup\tau_{1}}\frac{d\tau}{1+\frac{1}{2}\Phi(t)},\end{array}\label{eq: redshift 3}\end{equation}

which gives \begin{equation}
\int_{\tau_{1}}^{\tau_{0}+\bigtriangleup\tau_{0}}\frac{d\tau}{1+\frac{1}{2}\Phi(t)}=\int_{\tau_{0}}^{\tau_{1}+\bigtriangleup\tau_{1}}\frac{d\tau}{1+\frac{1}{2}\Phi(t)}.\label{eq: redshift 4}\end{equation}

This equation can be semplified, obtaining \cite{key-7}\begin{equation}
\int_{0}^{\bigtriangleup\tau_{0}}\frac{d\tau}{1+\frac{1}{2}\Phi(t)}=\int_{0}^{\bigtriangleup\tau_{1}}\frac{d\tau}{1+\frac{1}{2}\Phi(t)},\label{eq: redshift 4a}\end{equation}

which gives\begin{equation}
\frac{\bigtriangleup\tau_{0}}{1+\frac{1}{2}\Phi(t_{0})}=\frac{\bigtriangleup\tau_{1}}{1+\frac{1}{2}\Phi(t_{1})}.\label{eq: redshift 5}\end{equation}

Then \begin{equation}
\frac{\bigtriangleup\tau_{1}}{\bigtriangleup\tau_{0}}=\frac{1+\frac{1}{2}\Phi(t_{1})}{1+\frac{1}{2}\Phi(t_{0})}=\frac{a(t_{1})}{a(t_{0})}.\label{eq: redshift 6}\end{equation}

But frequencies are inversely proportional to times, thus \begin{equation}
\frac{f_{0}}{f_{1}}=\frac{\bigtriangleup\tau_{1}}{\bigtriangleup\tau_{0}}=\frac{1+\frac{1}{2}\Phi(t_{1})}{1+\frac{1}{2}\Phi(t_{0})}=\frac{a(t_{1})}{a(t_{0})}.\label{eq: redshift 7}\end{equation}

If one recalls the definition of the \textit{redshift parameter} \cite{key-16,key-17,key-18,key-19,key-20}
\begin{equation}
z\equiv\frac{f_{1}-f_{0}}{f_{0}}=\frac{\bigtriangleup\tau_{0}-\bigtriangleup\tau_{1}}{\bigtriangleup\tau_{1}},\label{eq: z}\end{equation}

using equation (\ref{eq: redshift 6}), equation (\ref{eq: z}) gives 

\begin{equation}
z=\frac{a(t_{0})}{a(t_{1})}-1,\label{eq: z 2}\end{equation}

which is well known in the literature \cite{key-16,key-17,key-18,key-19,key-20}.

Thus, it has been shown that the model is fine-tuned with the Hubble
Law and the Cosmological Redshift, exactly like the previous model
in \cite{key-7}.

\section{Conclusions }

An oscillating, homogeneous and isotropic Universe which arises by
the linearized Scalar-Tensor gravity has been discussed, integrating
my previous research in \cite{key-7} and showing that some observative
evidences, like the Cosmological Redshift and the Hubble law, are
fine-tuned with the model in this case too. In this context Dark Energy
appears like a pure curvature effect arising by the scalar field.

\section{Acknowledgements }

I would like to thank Herman Mosquera Cuesta for useful discussions
on the topics of this paper. The EGO consortium has also to be thanked
for the use of computing facilities

\end{document}